\documentstyle[epsfig]{aipproc}
\newcommand{\etal}{et~al.\/}

\newcommand{\HII}{\mbox {H\thinspace{\footnotesize II}}}

\newcommand{\Msun}{\mbox{${\cal M}_\odot$}}

\begin{document}

\title{Calibrating UV Emissivity And Dust Absorption 
At $z \approx 3$}
\author{Gerhardt R. Meurer$^*$, Timothy M. Heckman$^*$, and  
Daniela Calzetti$^{\dagger}$}
\address{$^*$The Johns Hopkins University, Baltimore, MD  21218\\
$^{\dagger}$Space Telescope Science Institute, Baltimore, MD 21218}

\maketitle

\begin{abstract}
We detail a technique for estimating the UV extinction and luminosity
of UV selected galaxies using UV quantities alone. The technique is
based on a tight correlation between the ratios of far infrared (FIR)
to UV flux ratios and UV color for a sample of local starbursts. A
simple empirical fit to this correlation can be used to estimate UV
extinction as a function of color.  This method is applied to a sample
of Lyman-break systems selected from the HDF and having $z \approx
3$. The resultant UV emissivity is at least nine times higher than the
original Madau \etal\ \cite{M96} estimate. This technique can be
readily applied to other rest-frame UV surveys.
\end{abstract}

\section*{Introduction}

Most of the light from high mass stars is emitted in the ultraviolet (UV;
$\lambda \approx 1100 - 3000$\AA), making it an attractive passband for
tracing cosmic star formation evolution.  This utility is accentuated with
increasing redshift as the rest-frame UV emission enters the optical where
modern detectors have quantum efficiencies approaching unity.  Unfortunately,
star formation occurs in a dusty environment, and dust efficiently absorbs
and scatters UV radiation.  This must also be the case in the early universe
since dust has been observed in objects with $z > 4$ (e.g. \cite{G97}).

The challenge of interpreting rest-frame UV emissivities is to devise an
adequate prescription to account for dust absorption.  Currently there is
much debate in the literature on what the proper dust correction prescription
is, resulting in different groups estimating $\lambda = 1600$\AA\ dust
absorption factors ranging from a factor of about 3 (e.g.\ \cite{P98}) to 20
\cite{SY97} at $z \approx 3$. The amount of high-$z$ dust absorption has a
direct bearing on interpretting how galaxies evolve.  Small dust corrections
favor hierarchical models of galaxy formation, while large corrections favor
monolithic collapse models \cite{MPD98}.

Here we consider the UV luminosity density at $z \approx 3$ derived mainly
from the $U$-dropouts in the Hubble Deep Field (HDF) \cite{W96}.  Our
technique \cite{M98} is based on the strong similarity between local
starburst galaxies and Lyman-break systems (e.g.\ \cite{L97}).  Throughout
this paper we adopt $H_0 = 50\, {\rm km\, s^{-1}\, Mpc^{-1}}$, $q_0 = 0.5$.

\section*{Method}

In earlier works \cite{M95,M97}, we showed that for local UV selected
starburst galaxies, the ratio of far infrared (FIR) to UV fluxes
correlates with UV spectral slope $\beta$ ($f_\lambda \propto
\lambda^\beta$ - $\beta$ is essentially an ultraviolet color).  This
is illustrated in Fig.~1.  Since $F_{\rm FIR}$ is dust reprocessed UV
flux, {\em this empirical correlation can be used to recover the
intrinsic UV flux from UV quantities alone.}  In addition, for
starbursts, the $y$ axis can be transformed directly into a UV
absorption \cite{M98}.  The fitted line is a simple linear fit to the
transformed data of the form: $A_{1600} \propto \beta - \beta_0$.

\begin{figure} 
\centerline{\epsfig{file=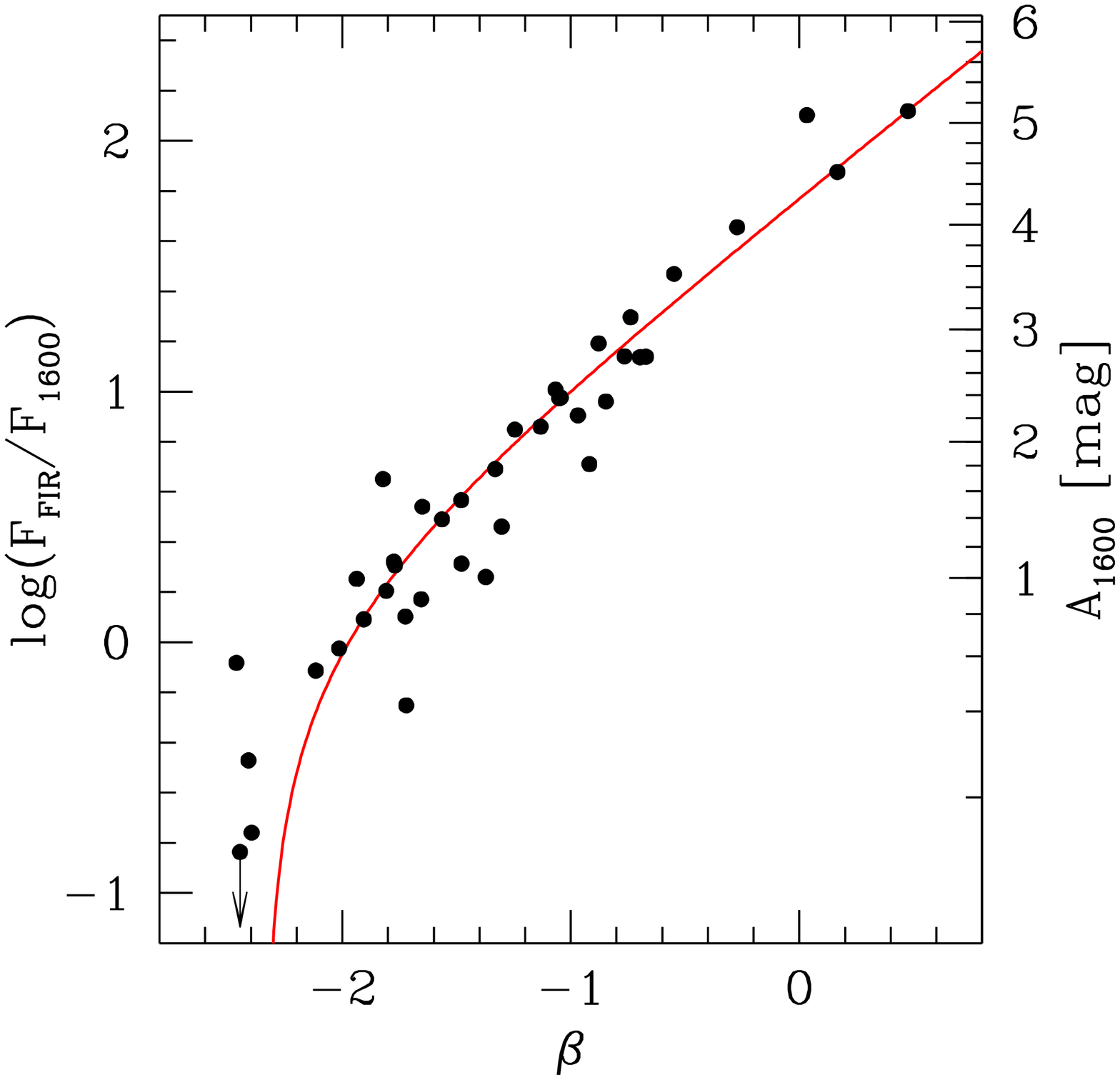,width=7.2cm,height=7.2cm}
\epsfig{file=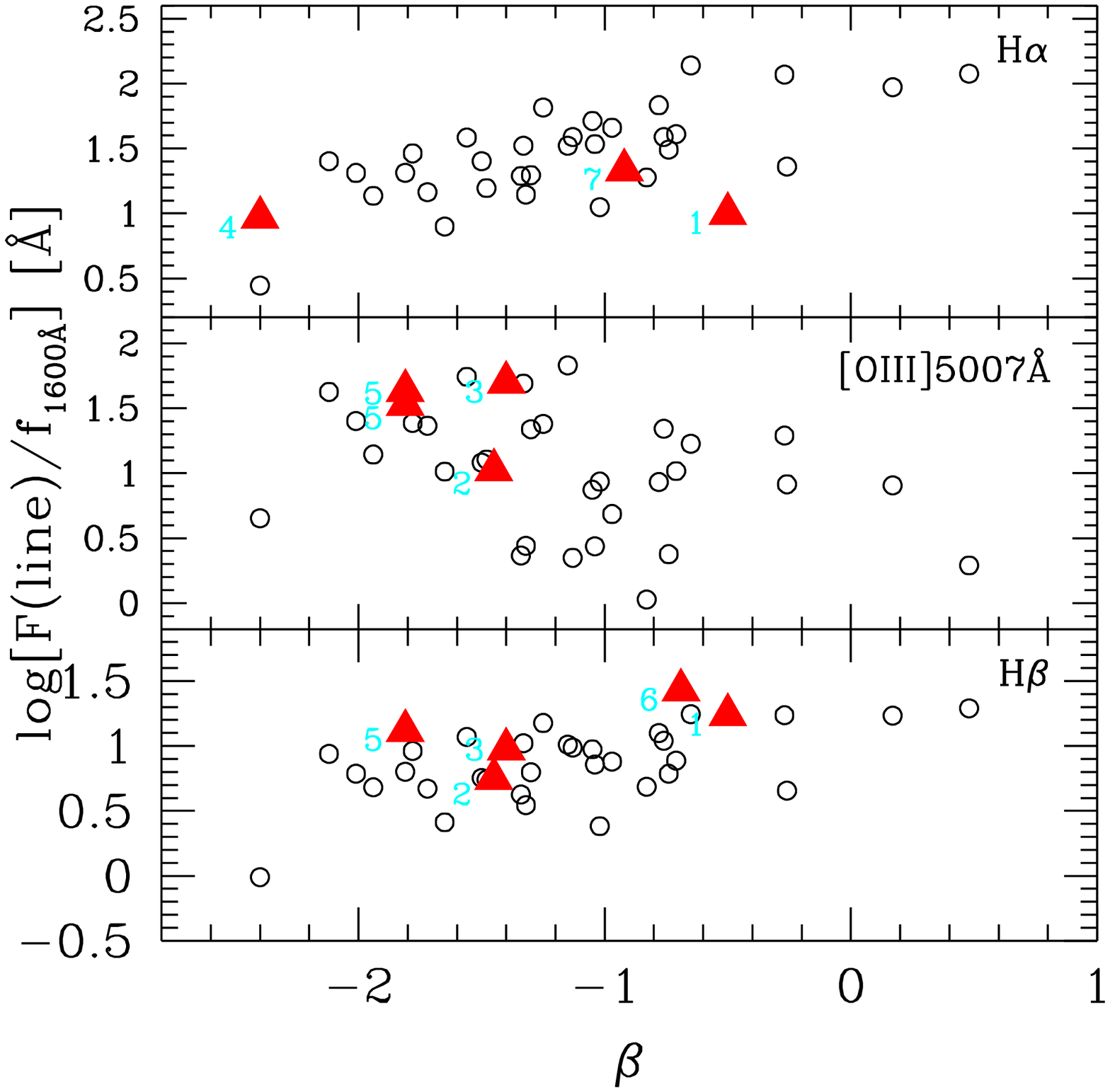,width=7.2cm,height=7.2cm}}
\vspace{10pt} 
{\small FIGURE 1 (left). FIR to UV flux ratio compared
to ultraviolet spectral slope $\beta$ for local UV selected
starbursts.\\ 
FIGURE 2 (right). Ratio of emission line flux to UV flux
density for local starbursts (circles), and HDF $U$-dropouts (numbered
triangles). }
\end{figure}

We selected our HDF $U$-dropout sample from a corner cut out of the $U_{300}
- B_{450}$ versus $V_{606} - I_{814}$ color-color diagram ($V_{606} - I_{814}
< 0.5$; $U_{300} - B_{450} \geq 1.3$) and adopted the same magnitude limits
as Madau \etal\ \cite{M96}.  We select in $V_{606} - I_{814}$ instead of
$B_{450} - I_{814}$ \cite{M96} because (1) $V_{606}$ is less affected by the
Lyman forest and edge than $B_{450}$, and (2) this selection yields fairly
even cutoff in $\beta$, and hence in $A_{1600}$.  Note that our selection
recovers high-$z$ galaxies in the ``clipped corner'' of the Madau \etal\
\cite{M96} selection area, and includes no known low-$z$ interlopers.

We applied our absorption law fit to broad-band $V_{606} - I_{814}$
colors transformed into $\beta$.  The transformation was derived from
high-quality IUE spectra that were ``redshifted'' through the $z = 2$ to
4 range of $U$-dropouts.  The transformation is linear in color with a
quadratic $z$ correction.  The $z$ correction is needed to account for
the Lyman forest and Lyman-edge creeping into the $V_{606}$ band at
high-$z$.

Figure~2 shows a test of our technique.  It compares the ratio of
(rest frame) optical emission line flux to UV continuum flux density
for local starbursts, and seven $U$-dropouts \cite{P98,WP98,B97,B98}.
These ratios are not corrected for dust absorption.  The overlap of
the two samples indicates that $U$-dropouts are ionizing populations
to the same degree as local starbursts.  Hence their intrinsic UV
spectrum should be similar.  Pettini \etal\ \cite{P98} claim that
$U$-dropouts probably suffer from little dust absorption since they
tend to have fairly low $F_{\rm H\beta} / f_{1600}$ values.  However,
this ratio can be misleading.  In fact, $F({\rm line})/f_{1600}$ is
not a good indicator of dust absorption: it does not correlate
strongly with $\beta$, which we know to be a good indicator of dust
absorption (Fig.~1).  This is the case for both the local and
$U$-dropout samples.  The reason for this was first proposed by
Fanelli \etal\ \cite{FOT88}: \HII\ emission lines are seen through a
larger column of dust than than the general UV continuum thus
cancelling the expected benefit in opacity of observing in the optical
instead of the UV.

\section*{Results}

Figure~3 plots the absorption corrected absolute AB magnitude of the
HDF $U$-dropouts versus $\beta$.  The broken lines show $M_{\rm
1600\AA}$ in the absence of absorption correction.  The data show an
apparent color - luminosity correlation.  This is in part due to the
selection limits, but the lack of very luminous blue galaxies is real.
This implies that there is a mass - metallicity relationship at $z
\approx 3$.  It also shows that the most luminous galaxies tend to
have the most dust absorption.  A similar color - luminosity
correlation is seen in local starbursts \cite{H98}.

\begin{figure} 
\centerline{\epsfig{file=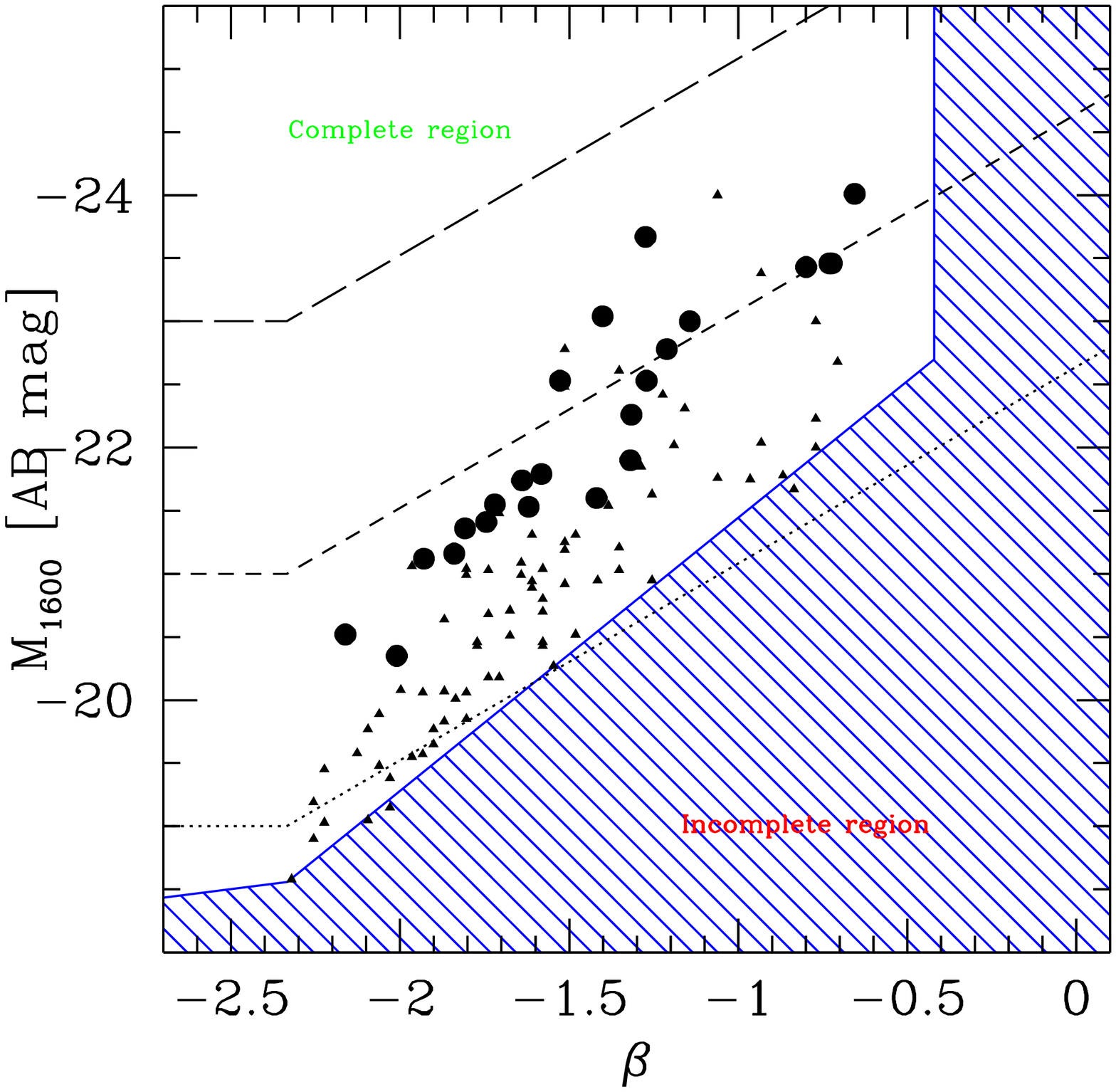,width=7.2cm,height=7.2cm}}
\vspace{10pt}
\centerline{\small FIGURE 3. Color - absolute magnitude diagram of HDF $U$-dropouts.}
\end{figure}

Summing the results for the HDF $U$-dropouts yields {\em lower limits}
to the intrinsic UV  emissivity, and hence the star formation rate density:

$$ \rho_{1600,0}  \gtrsim  1.5 \times 10^{27}\, {\rm erg\, s^{-1}\, Hz^{-1} Mpc^{-3}}$$
$$ \rho_{\rm SFR} \gtrsim 0.19\, \Msun\, {\rm yr^{-1}\, Mpc^{-3}} $$

We find that $\rho_{1600,0}$ is factor of 9.2 higher than $\rho_{1600}$
first estimated by Madau \etal\ \cite{M96}.  This difference is due to two
effects: the dust absorption correction (factor of 5.5), and the
improved $U$-dropout selection (factor of 1.7). These emissivities are
still lower limits because we have made no completeness corrections, and
because our $V_{606} - I_{814}$ selection is only sensitive to galaxies
with $A_{1600} \lesssim 3.4$ mag.  

Recently, Madau \etal\ \cite{MPD98} (see also Madau, this volume) have
fit models to cosmological emissivity data covering rest-wavelengths
from the FIR to the UV and redshifts out to $\sim 4$.  Their HDF
$U$-dropout sample now has a selection similar to ours.  Our
$\rho_{1600,0}$ estimate for the $U$-dropouts is a factor of 2.5 larger
than their preferred model, which simulates the heirarchical collapse
scenario and which includes a small amount of dust absorption.  However
it is only 30\%\ larger than their ``monolithic collapse'' model.
Hence, the initial phase of galaxy collapse was probably more rapid than
predicted by heirarchical models and somewhat obscured from our view by
at least modest amounts of dust.

%
%
%
%
%
%
%
%

\end{document}